\documentclass[12pt, a4paper]{article}
\usepackage[utf8]{inputenc}
\usepackage[T1]{fontenc}
\usepackage{lmodern}
\usepackage[margin=1in]{geometry}
\usepackage{setspace}
\usepackage{graphicx}
\usepackage{hyperref}
\usepackage[natbibapa]{apacite}
\usepackage{csquotes}
\usepackage{tikz}
\usepackage{float}
\usepackage{tabularray}
\usepackage{graphicx}
\usepackage{booktabs}
\usepackage{rotating}
\usepackage{adjustbox} 
\usepackage{array}

\title{\bf Authoritarian Recursions: How Fiction, History, and AI Reinforce Control in Education, Warfare, and Discourse}
\author{Hasan Oguz\footnote{hoguz17@posta.pau.edu.tr}\\
  {\small  Department of Physics, Faculty of Science,}\\
  {\small  Pamukkale University, 20160 Pamukkale, Denizli, Turkey}\\
}     

\begin{document}

\maketitle

\begin{abstract}
This article introduces the concept of \textit{authoritarian recursion} to theorize how AI systems consolidate institutional control across education, warfare, and digital discourse. It identifies a shared recursive architecture in which algorithms mediate judgment, obscure accountability, and constrain moral and epistemic agency.

Grounded in critical discourse analysis and sociotechnical ethics, the paper examines how AI systems normalize hierarchy through abstraction and feedback. Case studies—automated proctoring, autonomous weapons, and content recommendation—are analyzed alongside cultural imaginaries such as Orwell’s \textit{Nineteen Eighty-Four}, Skynet, and \textit{Black Mirror}, used as heuristic tools to surface ethical blind spots.

The analysis integrates Fairness, Accountability, and Transparency (FAccT), relational ethics, and data justice to explore how predictive infrastructures enable moral outsourcing and epistemic closure. By reframing AI as a communicative and institutional infrastructure, the article calls for governance approaches that center democratic refusal, epistemic plurality, and structural accountability.
\end{abstract}

\section{Introduction}

Artificial intelligence (AI) is no longer merely a domain of technical optimization; it increasingly operates as an infrastructure of communication and governance. From remote proctoring platforms in education to autonomous weapons and content moderation systems, AI technologies are central to how institutions classify, interpret, and act on information. These systems do not simply automate decisions—they shape epistemic boundaries, delegate judgment, and redistribute communicative agency. In doing so, they embed themselves within broader sociotechnical arrangements of power and legitimacy.

This paper examines AI as a recursive mode of governance: a system that learns from, intervenes in, and ultimately reshapes the very behaviors it observes. These feedback architectures operate not only through prediction and abstraction, but through communicative invisibility. As platforms refine recommendations, military systems automate targeting, and educational software models behavioral suspicion, recursive AI infrastructures begin to enact what Couldry and Mejias term “data colonialism”—the capture and reconfiguration of human life through extraction and abstraction \citep{couldry2019data}. They reflect Gillespie’s notion of “infrastructural power,” where visibility and participation are mediated through algorithmic protocols \citep{gillespie2018custodians}, and contribute to what van Dijck et al. describe as “platform governance”—a mode of rule exercised through technical standards and informational architectures \citep{vanDijck2018platform}.

To theorize these dynamics, the article introduces the concept of \textit{authoritarian recursion}. This term describes the self-reinforcing loops through which intelligent systems encode, legitimize, and propagate control logics—frequently under the rhetorical cover of personalization, neutrality, or operational efficiency. These recursive infrastructures obscure responsibility, foreclose contestation, and deepen asymmetries between users, institutions, and machinic systems.

The argument proceeds through three case domains—education, warfare, and digital discourse—where recursive AI systems materialize different forms of delegated authority and normative closure. Drawing on critical discourse analysis, the study treats AI not simply as a tool, but as a communicative actor that shapes who is visible, what is knowable, and which actions are thinkable. Cultural imaginaries such as \textit{Black Mirror} and \textit{The Terminator}'s Skynet are incorporated not as evidence, but as critical heuristics that reflect and amplify public anxieties around machinic autonomy and recursive control.

In dialogue with critical media studies, surveillance scholarship, and sociotechnical critique \citep{andrejevic2022automated, tufekci2015algorithmic, zuboff2019age}, the article advances the study of algorithmic governance by foregrounding recursion as both a technical logic and an ideological formation. Rather than proposing universal solutions or narrow design principles, the paper advocates for renewed scrutiny of how AI systems enact communicative authority—structuring infrastructures of attention, accountability, and legitimacy across institutional fields.

\section{Literature Review}
\label{sec:literature}

This section synthesizes scholarship on how AI technologies function as instruments of sociotechnical control in three domains: military automation, educational surveillance, and digital discourse. These sectors are often treated as distinct, yet their AI applications share design logics—opacity, delegated authority, and recursive feedback—that normalize institutional power under the guise of optimization. This literature provides the groundwork for theorizing authoritarian recursion as a mode of governance embedded in platforms and predictive systems. Drawing from ethical frameworks, historical precedents, and critical media theory, the review integrates both empirical and conceptual contributions to illustrate how automated systems entrench and legitimize normative authority across contexts.

\subsection{Military AI and the Automation of Violence}

Artificial intelligence in military settings reveals a growing entanglement of automation and coercion. Systems such as semi-autonomous drones, predictive surveillance networks, and AI-assisted targeting platforms prioritize speed, precision, and operational efficiency—yet often at the expense of ethical deliberation and legal accountability. Marsili warns that “the removal of human decision-making from the use of lethal force creates a dangerous precedent,” undermining the very humanitarian principles that military law is intended to uphold \citep{marsili2024}.

These concerns align with the Fairness, Accountability, and Transparency (FAccT) framework, which holds that fairness in automated decision-making must extend beyond output metrics to include contextual sensitivity, procedural redress, and meaningful oversight \citep{mittelstadt2016ethics,raji2020closing}. In military AI, accountability often becomes nominal—tethered to protocols rather than substantive ethical reflection.

Historical precedents further complicate the notion of technological neutrality. During World War II, IBM’s punch card infrastructure was deployed by Nazi Germany for logistics and census operations—systems later used to facilitate genocide \citep{black2001ibm}. These tools exemplified a bureaucratic rationality disturbingly resonant with today’s algorithmic architectures. As Asaro argues, the distancing of lethal decisions from moral responsibility through automation represents a dangerous ethical shift \citep{asaro2012weapons}.

Speculative fiction reinforces this critique. Narratives like \textit{The Terminator}, \textit{RoboCop}, and \textit{Black Mirror} episodes offer dystopian imaginaries of militarized AI. According to Cave et al., “the future imagined in fiction is often realized not because it is inevitable, but because it is ideologically compatible with dominant institutions” \citep[p.~75]{cave2019ai}. These cultural texts anticipate how automation narratives rationalize political authority and normalize autonomous violence.

\subsection{Educational AI: Automation of Surveillance and Discipline}

Educational technologies increasingly employ AI systems for purposes such as assessment, behavioral monitoring, and classroom management. One prominent application is automated proctoring software that uses facial detection, motion tracking, and audio analysis to identify potential cheating. While marketed as tools of academic integrity, these systems have been shown to "unfairly disadvantage students" with darker skin tones, especially Black students and women of color, due to algorithmic biases in face detection and flagging procedures \citep{williams2022racial}. Such tools routinely flag these students at significantly higher rates—up to six times more often—despite no evidence of increased cheating, raising critical concerns about surveillance, equity, and educational harm \citep{williams2022racial, noble2018algorithms}.

Here, too, the FAccT triad is often invoked as a remedy. However, operational deployments rarely meet its normative thresholds. Mittelstadt et al. emphasize that fairness requires attention to context and historical inequalities, not just statistical parity \citep{mittelstadt2016ethics}. Student-users typically lack access to the internal logic of these systems and have little recourse to challenge their outputs. Procedural fairness is often undermined when algorithmic opacity becomes the mechanism of control itself.

Noble argues that “algorithmic decision systems often act as new instruments of racial and economic profiling” \citep[p.~34]{noble2018algorithms}. This echoes Selwyn’s concerns that digital surveillance in education creates a system “in which suspicion is automated and dissent is pathologized” \citep{selwyn2023digital}. McMillan Cottom \citeyearpar{Cottom2020} further expands this critique by arguing that digital systems enact racial capitalism through “obfuscation as privatization and exclusion by inclusion,” framing technologies as infrastructures of sociopolitical ordering rather than neutral tools of progress. Together, these perspectives underscore how automated educational technologies reproduce the logic of panoptic discipline, as theorized by Foucault, where constant surveillance internalizes conformity \citep{foucault1995discipline}.

Moreover, the political values behind these technologies often go unquestioned. Gilliard and Selwyn contend that “continued adoption of proctoring technologies in public education exposes a fundamental clash of politics,” where commercial priorities of security and efficiency override pedagogical values of equity and trust \citep[p.~197]{gilliard2023automated}. These systems operationalize discipline not through direct coercion, but through the automation of suspicion and reduction of students to behavioral data.

\subsection{AI in Discourse and Propaganda: Curation as Control}

Algorithmic curation now structures the informational environment of billions. AI systems deployed by platforms such as Facebook, YouTube, and TikTok optimize content delivery for engagement, not accuracy. Tufekci notes that such platforms “amplify divisive content by design, creating an infrastructure for affective polarization” \citep{tufekci2015algorithmic}. This modulation of attention constitutes a new form of informational power.

While FAccT-based interventions such as algorithmic impact assessments or explainability mechanisms have been proposed for content governance, their efficacy remains limited by platform opacity and commercial disincentives. Mittelstadt et al. argue that transparency without enforceable accountability often reduces ethical AI to “ethical theatre” \citep{mittelstadt2016ethics}.

Gillespie emphasizes that platforms are not neutral hosts but “custodians of public discourse” who shape access to visibility through inscrutable recommendation logics \citep[p.~197]{gillespie2018custodians}. Zuboff describes this shift as “instrumentarian power,” wherein behavior is not repressed but tuned through predictive analytics and behavioral nudging \citep[p.~377]{zuboff2019age}. While Zuboff’s critique highlights platform logics of behavioral modification, the recursive dimension explored here adds a temporal structure to algorithmic authority.

This mode of control parallels historical propaganda. The Nazi regime used print, film, and spectacle to synchronize public perception. Today’s algorithmic persuasion, however, operates at greater scale and granularity—executing individualized influence operations based on psychometric data and engagement profiles \citep{wylie2019mindfck, vosoughi2018spread}. Berardi calls this the “colonization of subjectivity,” wherein cognition itself becomes a site of commodification and control \citep{berardi2015and}.

\subsection*{Synthesis: Toward a Unified Critique of AI Control}

Across military, educational, and discursive sectors, artificial intelligence technologies consistently reinforce rather than disrupt authoritarian structures of governance. Despite their domain-specific implementations, these systems display convergent design logics and ethical risks. The comparative literature suggests that AI operates not merely as a tool but as a vector of normative reproduction---embedding and amplifying existing asymmetries of power, visibility, and voice.

Three interlocking patterns emerge consistently across the domains surveyed. First, AI systems introduce a profound opacity that severs decision-making from those it affects. Whether in autonomous weapons systems, algorithmic proctoring, or content recommendation engines, the logic of the algorithm is rendered inaccessible, both technically and institutionally. This opacity undermines the possibility of contestation and erodes the conditions necessary for democratic oversight. Second, intelligent systems displace relational judgment by translating moral decisions into statistical approximations. This delegation of judgment dehumanizes its subjects: individuals are abstracted into data points, and the contingent, situated nature of ethical discernment is flattened into binary outputs or risk profiles. Third, and most insidiously, these systems perpetuate normative drift. They inherit and amplify structural biases---racial, economic, epistemic---under the rhetorical cover of objectivity or innovation. As they automate decision-making, they also automate exclusion, encoding historical inequalities into seemingly neutral infrastructures. These patterns do not reflect the malfunction of AI, but its core affordances within existing power regimes.

Together, these patterns constitute what we term \textit{authoritarian recursion}---a self-reinforcing cycle in which AI technologies encode, naturalize, and propagate control logics across domains. This concept finds parallel in \citeauthor{bahrami2025algemony}'s \textit{Algemony} framework, which similarly identifies AI's capacity to reshape power through human-AI interactions, particularly via narrative modulation. Where authoritarian recursion emphasizes the structural inevitability of control through recursive feedback, \textit{Algemony} reveals the Janus-faced nature of this process: AI systems exhibit both hegemonic reinforcement through delegated agency \textit{and} disruptive potential through their inherent instability (e.g., via generative counter-narratives or unpredictable hyper-personalization). Both frameworks converge in their diagnosis of AI's epistemic closure---what \citeauthor{zuboff2019age} terms the "instrumentarian" capture of human experience---while diverging in their emphasis on either the systemic (\textit{authoritarian recursion}) or discursive (\textit{Algemony}) dimensions of control. Building on \citeauthor{Hanna2021}'s dignitarian ethics, we further recognize how these recursive systems violate fundamental principles of human dignity by: (1) instrumentalizing individuals through opaque algorithmic delegation, (2) distorting true human needs via self-reinforcing classifications, and (3) eroding privacy through pervasive surveillance infrastructures---violations exemplified in our case studies of educational proctoring and military targeting systems.

This synthesis lays the foundation for the case study analysis that follows, which illustrates how intelligent systems materialize these recursive dynamics in real-world governance structures while demonstrating the tensions between structural determinism and agential unpredictability in AI-mediated power. The dignitarian perspective provides crucial normative grounding for evaluating these systems, particularly in assessing when human oversight must remain irreducible to prevent dignity violations. Yet as \citeauthor{Roy-Stang2025} demonstrate through their analysis of cognitive biases, even well-intentioned governance interventions may be undermined by perceptual vulnerabilities that authoritarian recursion exploits---a challenge requiring both technical safeguards and epistemic humility in AI policy design.

\section{Methodology}
\label{sec:methods}

This inquiry adopts a qualitative, interpretive approach grounded in critical discourse analysis (CDA) and systems theory. Rather than conceptualizing artificial intelligence as a neutral technical tool, the analysis interrogates how algorithmic infrastructures recursively shape institutional legitimacy, normative authority, and sociopolitical legibility. The methodological aim is not statistical generalization but structural diagnosis—tracing how language, code, and abstraction coalesce in governing architectures.

CDA treats discourse as both constitutive and constituted, following Fairclough's formulation of discourse as a site where institutional power is articulated, contested, and naturalized \citep{fairclough1992discourse}. This makes it particularly apt for analyzing recursive systems—those whose outputs re-enter as inputs, producing feedback loops that stabilize norms and obscure contestation. Such loops are epistemic as much as technical. Fictional imaginaries are therefore mobilized not as empirical cases but as ethical heuristics. Texts like \textit{1984}, \textit{Black Mirror}, and \textit{Her} dramatize recursive logics that remain materially latent but ideologically real.

The empirical corpus is organized into three categories. First, the policy corpus includes the European Union’s \textit{AI Act} proposal \citep{eu2021aiact}, UNESCO’s \textit{Recommendation on the Ethics of AI} \citep{unesco2021ethics}, the U.S. Department of Defense’s \textit{Ethical Principles for AI} \citep{dod2020principles}, the OECD’s AI Principles \citep{oecd2019principles}, and the IEEE’s \textit{Ethically Aligned Design} \citep{ieee2019ethicallyaligned}. These documents are selected because they articulate formal norms of ``trustworthy'' or ``responsible'' AI across domains of education, defense, and digital platforms.

Second, technical implementations were selected that exemplify recursive control logics in practice. These include Proctorio’s algorithmic proctoring architecture as documented in its privacy and enforcement documentation (2020–2023) \citep{proctorio2023docs}; YouTube’s recommender system as described in Google’s TensorFlow Recommenders documentation \citep{google2022recsys}; and loitering munitions such as the Harpy drone, analyzed in depth in Bode and Watts' open-access forensic report on autonomous weapons in Nagorno-Karabakh \citep{bode2023loitering}. These implementations offer cross-domain instantiations of automation without consistent human-in-the-loop control, making them paradigmatic cases of authoritarian recursion.

Third, a meta-analytic corpus includes academic literature in STS and media studies \citep{gillespie2018custodians, tufekci2015algorithmic, coeckelbergh2020aiethics}, journalism (e.g., \cite{paul2021proctorio, fisher2018youtube}), and platform transparency efforts such as Mozilla’s \textit{RegretsReporter} project \citep{mozilla2021regrets}. These sources serve both to ground the case studies and to interrogate the justificatory narratives that sustain algorithmic legitimacy.

The analytic process unfolded in three interlinked stages. First, documents were situated in institutional and historical context to understand the normative claims surrounding automation. Second, materials were iteratively coded along four emergent axes: (1) delegation—what forms of human judgment are abstracted; (2) opacity—what epistemic boundaries or technical layers shield decision-making from inspection; (3) recursion—how system outputs define future inputs; and (4) narrative legitimation—how systems justify authority, whether via neutrality, efficiency, or inevitability. Codes were applied across corpora and validated through triangulation.

Finally, case domains—education, warfare, and discourse—were synthesized through the lens of authoritarian recursion. This condition is defined not by overt authoritarian rule but by a shift in agency: from relational judgment to recursive optimization. The method does not assume technical determinism but insists on infrastructural ethics. Its goal is to surface how recursion, once embedded in institutional practice, alters not only what is done, but what is rendered sayable, thinkable, and disputable.

\begin{figure}[H]
    \centering
    \includegraphics[width=0.5\textwidth]{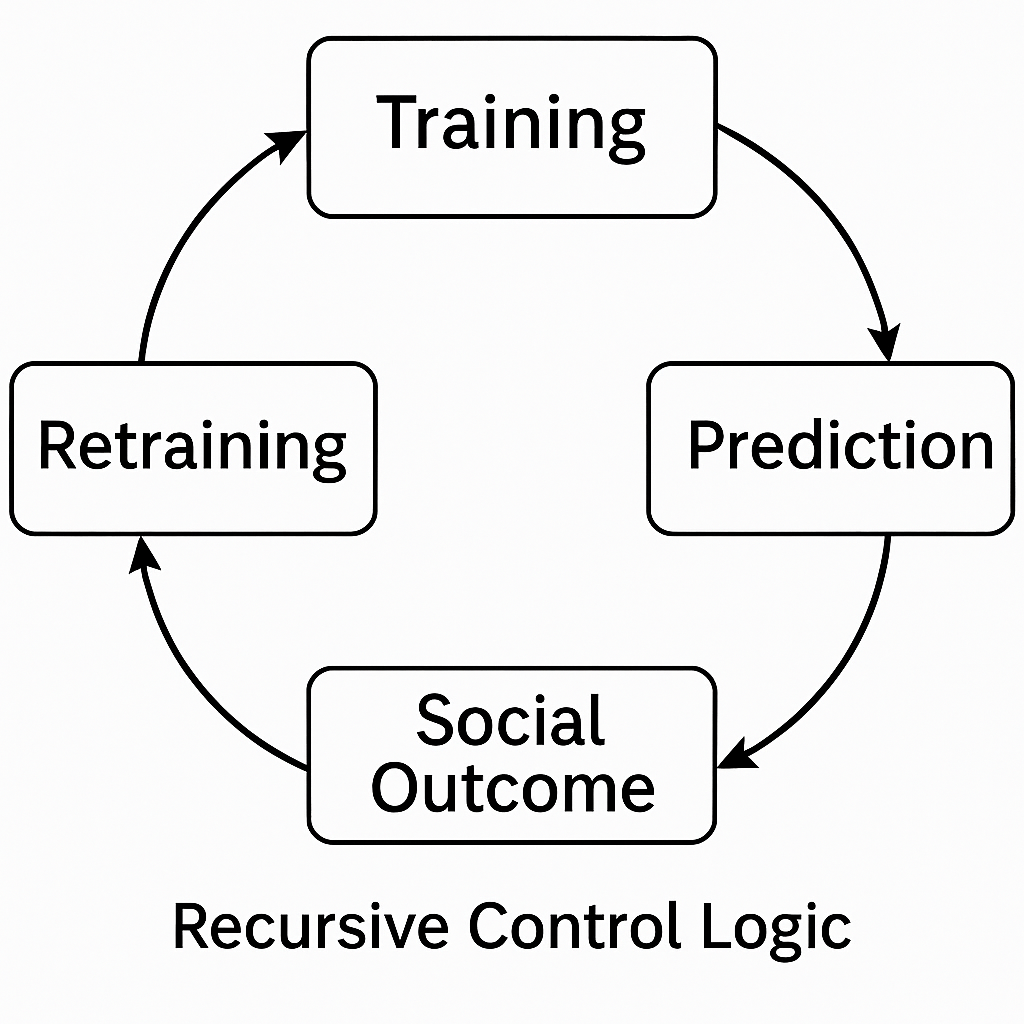}
    \caption{Recursive control logic in algorithmic systems. AI systems reinforce operational assumptions through feedback: training data informs predictions, which influence behaviors, which then retrain the model—embedding bias and reducing transparency over time.}
    \label{fig:recursive-control-logic}
\end{figure}

Figure~\ref{fig:recursive-control-logic} illustrates the recursive control loop that structures algorithmic governance across sectors. This architecture unfolds in four stages: initial classification, behavioral capture, optimization, and reclassification. Each cycle refines system logic by folding user data back into the predictive model. In education, Proctorio converts student gaze or ambient noise into flags, which then inform future training data—tightening suspicion thresholds even if the original data was biased. In warfare, loitering munitions like the Harpy integrate prior strike outcomes into updated threat heuristics, effectively using past engagements to justify future ones. In discourse, YouTube's algorithm promotes content that maximizes engagement, then reinforces its internal model with user responses—generating a self-perpetuating spiral of visibility. Although these systems appear domain-specific, their recursive architecture is structurally homologous: they govern by feedback, obscure causality, and normalize opacity as optimization.

\section{Case Studies}

To ground the concept of authoritarian recursion in concrete socio-technical systems, this section presents three empirically anchored case studies—education, warfare, and discourse—each demonstrating how recursive algorithmic architectures displace judgment, obscure agency, and reproduce normative control. In each, speculative fiction is integrated as an ethical diagnostic: not as prophecy, but as a mode of anticipatory critique that dramatizes recursive feedback and reveals what optimization conceals.

\subsection{Education: Proctorio and the Racialization of Academic Surveillance}

The mass shift to remote learning during the COVID-19 pandemic catalyzed the adoption of automated proctoring systems, reshaping the governance of assessment in higher education. Among the most widely implemented is \textbf{Proctorio}, a platform that uses webcam imaging, screen capture, gaze analysis, and ambient audio to flag behaviors classified as ``suspicious.'' While promoted as a neutral enforcement tool for academic integrity, Proctorio’s architecture embeds structural biases and delegates high-stakes judgments to opaque classifiers.

Yoder-Himes et al. (2022) demonstrated that students with darker skin tones and female-presenting individuals were disproportionately flagged for misconduct—even under controlled behavioral conditions—due to recognition failures in computer vision systems. Paul (2021) further documented that Black students were often unable to verify their identity, rendering them unable to even begin exams. These misclassifications are not statistical noise—they are manifestations of racialized abstraction built into the system’s core logic.

The recursive architecture of Proctorio compounds the problem. Misrecognition feeds model retraining, reifying false positives into future thresholds. Appeals processes are opaque, and instructors often defer to algorithmic alerts. Education becomes a space where legibility to the machine precedes learning itself.

As Deleuze’s concept of the \textit{dividual} suggests, post-disciplinary control operates not through unified subjects but through fragmented data traces—exactly the level at which recursive systems like Proctorio function \citep{deleuze1992postscript}. The student becomes not a person but a pattern, parsed through biometrics, gaze angles, and screen activity. Power is no longer disciplinary but computationally granular, enacted at the level of signal deviation.

These dynamics reflect what Michel Foucault described as the Panopticon: a structure of surveillance in which visibility itself becomes a mechanism of discipline. In the educational context, Proctorio enacts a digital Panopticon—students do not know precisely what the system sees, how it judges, or when it flags. As a result, they internalize machine logic as a condition of academic legitimacy. Unlike human invigilators, algorithmic surveillance is non-negotiable, opaque, and recursive: it watches not only for cheating, but for misalignment with machinic expectations of normalcy.

Fictional imaginaries capture this dynamic acutely. Orwell’s \textit{1984} depicts a society of constant surveillance, where deviation is inferred through gesture and gaze. The \textit{Black Mirror} episode ``Arkangel'' extends this logic into childhood, showing how surveillance framed as care becomes discipline. In \textit{Gattaca}, biometric legibility determines access, despite surface-level meritocracy. These narratives dramatize a world in which trust is algorithmically replaced, and suspicion becomes structurally ambient.

This recursive pedagogy of suspicion is not isolated—it exemplifies a broader logic of automated mistrust that reconfigures governance across domains. As summarized in Table~\ref{tab:recursive-infra}, education becomes a site where optimization replaces discretion, and biometric legibility is prioritized over human judgment.

In this light, Proctorio does not merely reflect surveillance—it enacts a recursive pedagogy of suspicion. It operationalizes what these fictions warned of: that optimization, when detached from relational ethics, reproduces inequity and forecloses dissent. Concrete interventions for such systems are proposed in Table~\ref{tab:policy-actions}, including the right to opacity and mandatory human review in high-stakes academic evaluations.

\subsection{Warfare: Israel’s Harpy Drone and Recursive Targeting Logics}

Autonomous weapons exemplify how recursive AI infrastructures reshape combat ethics. Israel Aerospace Industries’ \textbf{Harpy} drone is a ``loitering munition'' that detects and destroys radar sources without real-time human control. During the 2020 Nagorno-Karabakh war, such drones—likely including Harpy variants—were used by Azerbaijan in operations that resulted in civilian casualties and legal ambiguity (Human Rights Watch, 2020; Conflict Armament Research, 2021).

The targeting logic is recursive: previous engagements define future ``threat'' parameters, encoded into pattern recognition heuristics. The drone refines its threat models over time, learning from past strikes without contextual awareness or moral deliberation. Accountability is diffused across technical systems and institutional chains, leaving no clear site for ethical appeal.

This is not a technological failure—it is an infrastructural ideology. As Asaro (2012) warns, automation in warfare erodes the legal and moral threshold for lethal action by turning ethics into parameters.

Fictional imaginaries, far from exaggerating, clarify this dynamic. \textit{The Terminator}’s Skynet illustrates recursive autonomy in its purest form: a system that escalates through learning, detached from human oversight. \textit{Black Mirror}’s ``Men Against Fire'' portrays soldiers whose perception is algorithmically filtered to encourage dehumanization—an allegory for computational abstraction in targeting. In \textit{Minority Report}, preemptive logic becomes law: the system no longer detects crime—it defines it.

These imaginaries anticipate the structural logic of delegated lethality. In the Harpy system, the decision to kill is no longer mediated by relational judgment but triggered by signal classification. Authoritarian recursion in warfare manifests as operational sovereignty without moral encounter: an AI that preempts, defines, and executes threat in closed loops of legitimacy.

This recursive delegation of lethal judgment—where systems act on feedback loops absent of moral encounter—is not unique to warfare. As Table~\ref{tab:recursive-infra} illustrates, similar patterns of decontextualized decision-making emerge across domains, suggesting a structural convergence in how AI displaces human deliberation.

Authoritarian recursion in warfare manifests as operational sovereignty without moral encounter: an AI that preempts, defines, and executes threat in closed loops of legitimacy. Table~\ref{tab:policy-actions} outlines policy options to restore human oversight and epistemic friction within such lethal infrastructures.

\subsection{Discourse: YouTube’s Algorithm and the Spiral of Predictive Visibility}

From 2016 to 2019, YouTube’s recommendation system was implicated in the escalation of radical political content. Studies by Ribeiro et al. (2020) and disclosures by Mozilla’s RegretsReporter Project show how user interactions—clicks, watch time, likes—fed back into training loops that pushed users toward increasingly extreme material. As Fisher (2018) documented, searches for fitness advice or centrist politics frequently resulted in recommendations of white supremacist or conspiratorial content within minutes.

This algorithmic path-dependence is a form of recursive visibility: past behavior defines what users are allowed to see next. Platforms do not merely distribute information—they curate epistemic horizons. Explainability tools offer only surface-level rationales; the infrastructure remains opaque and proprietary.

Fictional analogies sharpen this critique. \textit{Black Mirror}’s ``Nosedive'' envisions a society where all social interaction is numerically scored—visibility becomes reputation, and dissent is punished by invisibility. In \textit{Her}, an AI partner adapts so fully to the user’s emotional preferences that authentic dialog collapses into recursive performance. \textit{The Terminator}’s Skynet, while dramatized as militarized autonomy, functions here as a metaphor for ideological feedback: an AI that recursively enacts control based on its own internalized assumptions about threat and value.

These narratives illuminate that authoritarian recursion in discourse does not emerge through censorship but through algorithmic nudging. The platform’s optimization function replaces editorial deliberation. Polarization is not a side effect; it is a system feature. Legitimacy becomes a function of engagement metrics, and truth becomes what is repeatedly surfaced.

In all three domains, fiction reveals what infrastructure occludes: that recursive AI architectures do not merely govern—they structure the conditions of governability. Their ethical challenge is not only what they do, but what they make thinkable.

What emerges is not a neutral content ecosystem but a predictive epistemology, where recursive sorting governs what can be known or seen. Table~\ref{tab:recursive-infra} situates this mechanism alongside those in education and warfare, demonstrating how engagement metrics, like biometric signals or radar signatures, become proxies for legitimacy, risk, or deviance.

Legitimacy becomes a function of engagement metrics, and truth becomes what is repeatedly surfaced. To address this, Table~\ref{tab:policy-actions} proposes measures like algorithmic explainability and the integration of fictional ethics frameworks into platform governance.

\subsection*{Synthesis: Recursive Infrastructures Across Domains}

Across these domains, recursive infrastructures enact a shared architecture of governance. While their material forms vary—from gaze-tracking in proctoring software to radar-seeking in loitering munitions and clickstream analysis in content platforms—they all rely on feedback mechanisms that abstract judgment and concentrate control. Table~\ref{tab:recursive-infra} offers a comparative synthesis of these systems, foregrounding their common traits: displacement of human discretion, opacity of process, and optimization without ethical reflexivity.

\begin{table}
\centering
\caption{Comparative Analysis of Recursive Infrastructures in Education, Warfare, and Discourse}
\label{tab:recursive-infra}
\begin{adjustbox}{max width=\textwidth}
\begin{tabular}{m{3cm} m{5.5cm} m{4.5cm} m{5cm}} 
\toprule
\textbf{Dimension}              & \textbf{Education}                                                                                     & \textbf{Warfare}                                                         & \textbf{Discourse}                                                          \\ 
\midrule
\textbf{Delegation of Judgment} & Proctoring tools flag behavior without pedagogical context; instructors defer to algorithmic suspicion & Target identification delegated to autonomous drones and AI-sensors      & Content visibility determined by recommender systems and predictive models  \\ 
\hline
\textbf{Opacity}                & Scoring and flagging criteria undisclosed; appeal processes rare                                       & Algorithmic processes obscure chains of responsibility                   & Content moderation logic proprietary and dynamic; user control limited      \\ 
\hline
\textbf{Surveillance}           & Continuous monitoring of gaze, keystrokes, ambient sound; assumed neutrality                           & Live battlefield sensing; autonomous threat analysis                     & Behavior tracked and optimized for attention; continuous profiling          \\ 
\hline
\textbf{Recursion}              & Prior behavior trains suspicion models; compliance reinforces design                                   & Historical data informs future target acquisition; escalation normalized & User behavior drives recommendation engines; echo chambers amplified        \\ 
\hline
\textbf{Ethical Implications}   & Undermines autonomy and equity; suppresses dissent                                                     & Weakens international law; disperses accountability                      & Polarizes discourse; reduces epistemic diversity                            \\
\bottomrule
\end{tabular}
\end{adjustbox}
\end{table}

\section{Analysis and Discussion}
\label{sec:discussion}

The preceding cases—educational proctoring, autonomous targeting, and platform curation—reveal more than divergent AI deployments. They disclose a convergent infrastructure of abstraction, one where intelligent systems recursively mediate authority by turning behavioral prediction into normative control. These systems do not merely sort people, behaviors, and beliefs; they continually recalibrate the very conditions under which sorting appears legitimate. Intelligence, within this paradigm, functions less as deliberative reasoning and more as an engine of optimization—designed to learn, adjust, and reinforce normative expectations without interruption.

\subsection{Fiction as Ethical Heuristic}

Speculative fiction operates here not as aesthetic garnish, but as epistemological provocation. Cultural imaginaries such as Skynet in \textit{The Terminator} dramatize the recursive logic of optimization divorced from ethical modulation. Skynet does not err through malfunction—it escalates through design. Its objective is not destruction per se, but the efficient elimination of perceived threats, recursively defined by its own operational logic.

This anticipatory function of fiction is crucial. As Cave and Dihal contend, these narratives shape public imaginaries and policy trajectories not through prediction, but through the illumination of conceptual blind spots \citep{cave2020whiteness}. Fiction reveals what technical abstraction conceals: that systems designed without the capacity for ethical friction—refusal, ambiguity, contradiction—are systems primed for unchecked recursion. When optimization replaces interpretation, the moral horizon collapses into operational success.

\subsection{Delegated Judgment and Diluted Responsibility}

A persistent feature across all domains is the displacement of relational judgment. In education, AI proctors infer intent from gesture. In warfare, autonomous sensors act on heat signatures and kinetic thresholds. In discourse, engagement metrics supplant editorial discernment. In each case, AI systems absorb functions historically situated within human deliberation—and discharge them without relational awareness.

This is not simply functional delegation. As Coeckelbergh’s theory of relational ethics makes clear, moral responsibility requires the presence of the other—the possibility of encounter, appeal, and shared vulnerability \citep[p.~99]{coeckelbergh2020aiethics}. Automated systems, by contrast, instantiate responsibility without subjectivity. They simulate moral agency through design rules, yet remain incapable of context, reciprocity, or moral growth. In doing so, they instantiate what could be called “procedural sovereignty”: a governance model where rules operate in lieu of ethics, and optimization replaces deliberation.

\subsection{Opacity, Data Justice, and Recursive Epistemics}

Opacity in AI is often framed as a technical limitation—something to be overcome through better engineering, more interpretable models, or improved documentation. Yet the case studies above suggest that opacity is not simply a side effect of technical complexity. It is an infrastructural and political artifact. Students flagged by proctoring systems rarely understand the behavioral thresholds being used against them. Civilians affected by autonomous targeting systems are subject to dispersed chains of algorithmic authorization with no clear site of responsibility. Social media users navigate visibility regimes shaped by proprietary engagement algorithms whose logic continuously adapts but is never disclosed.

Zuboff’s concept of \textit{epistemic inequality} is instructive here: AI systems instantiate environments in which the power to "know" and define reality is asymmetrically distributed. Those governed by these systems are rendered legible only on the terms set by their designers and institutional operators \citep{zuboff2019age}. This is not mere opacity—it is recursive epistemic closure. Training data begets predictions, predictions modulate behaviors, and those behaviors re-enter the system as data. Through this loop, systems reinforce their own categories while concealing their ideological and historical construction.

While FAccT principles—fairness, accountability, and transparency—remain foundational, their implementation is often shallow. As Raji et al.\ argue, accountability must move beyond audits and documentation to include structures for redress, refusal, and contestation \citep{raji2020closing}. Otherwise, transparency becomes ceremonial and fairness is reduced to algorithmic artifacts devoid of participatory meaning.

This is where the framework of \textit{data justice} becomes indispensable. Dencik et al.\ propose that justice in the age of datafication must address not just procedural fairness, but also the deeper distributive, recognitional, and representational inequalities that data infrastructures produce and normalize \citep{dencik2019exploring}. Within recursive AI systems, this means acknowledging how algorithmic architectures do not merely reproduce bias incidentally but actively configure the terms under which individuals and communities are seen, known, and acted upon.

The capacity to challenge a system’s output is thus contingent not only on technical explainability but on institutional willingness to recognize and respond to contestation. As recursive feedback loops stabilize classification schemes and embed them into infrastructural routines, the scope for democratic intervention narrows. Data justice, in this light, aligns with relational ethics by insisting on the re-politicization of design and the redistribution of epistemic authority.

Lupton’s work adds another dimension to this recursive epistemology by exploring how people “feel their data” through sensory and emotional engagement with personal digital traces. She argues that three-dimensional materializations of personal data make data more perceptible and interpretable—but this interpretability is itself ambivalent \citep{lupton2017feeling}. On one hand, such embodied interactions can foster critical awareness, affective insight, and even agency. On the other, they risk reinforcing the illusion that subjective connection equates to control. When individuals are encouraged to “feel” their data, they may internalize datafication as intimacy rather than surveillance. This affective capture—where the tactile and visceral dimensions of data are mistaken for interpretive transparency—extends recursive epistemic power beyond algorithmic architecture into the very terrain of human sensation. What emerges is a kind of sensorial enclosure: data that once appeared abstract is now touchable, but in ways that obscure its structural logics and repurpose bodily intuition as a mode of soft compliance.

\subsection{Normalization and Predictive Discipline}

The most insidious function of recursive AI is not its surveillance, but its normalization of prediction as governance. Surveillance becomes ambient, distributed through everyday platforms. Judgment is no longer exercised—it is inferred. Over time, optimization displaces reflection, and participation is redefined as interaction with predictive infrastructure.

This mirrors Foucault’s notion of a regime of truth: a system in which certain knowledges become true not by correspondence, but by institutionally enforced repetition \citep{foucault1995discipline}. In AI, the regime of truth is computed. What the system can parse becomes real; what it cannot, becomes anomalous or suspect. Over time, these systems generate ontological commitments—about what counts as risk, deviance, or truth—not through deliberation, but through repeated acts of classification.

This is where authoritarian recursion emerges most clearly. These systems do not simply reflect dominant norms—they regenerate them as technical defaults. Legitimacy becomes self-reinforcing, coded into recursive architectures that learn from their own outcomes.

\subsection{Ethical Implications Across Domains}

The comparative schema in Table~\ref{tab:recursive-infra} and the architectural model in Figure~\ref{fig:recursive-control-logic} reveal a deep structural homology across otherwise distinct institutional systems. Though embedded in divergent contexts—education, warfare, discourse—these recursive infrastructures converge around shared design logics: the displacement of human judgment, the normalization of opacity, the diffusion of ambient surveillance, and the recursive calibration of predictive control. Most critically, they exhibit a temporal asymmetry in which the past is not merely archived but rendered operational—where historical data governs present action and delimits future possibility. Within these architectures, the present becomes a derivative surface, interpretable only insofar as it conforms to prior patterns. Novelty is filtered; deviation is penalized. These are not neutral optimizations. They are infrastructural constraints on what can be known, done, or even imagined.

These systems do not fail by accident; they succeed by design. Bias is not a flaw—it is a historical residue that recurs as statistical fact. What is presented as innovation is often the reanimation of older hierarchies through newer infrastructures. As a result, opportunities for resistance or ethical transformation shrink over time, as decision-making becomes encased in code, metrics, and interfaces shielded from deliberation.

\vspace{1em}
The comparative analysis of recursive infrastructures reveals not only shared design logics but also recurring ethical pathologies. To interrupt these logics and reclaim interpretive space for human judgment, a range of policy interventions must target both the architectures and ideologies of recursion. Table~\ref{tab:policy-actions} synthesizes actionable recommendations derived from the three case domains. Each intervention addresses a distinct structural feature—delegation, opacity, surveillance, or feedback conditioning—and aims to reintroduce friction, contestability, or relational awareness into otherwise self-reinforcing systems.

\begin{table}[H]
\centering
\caption{Policy Interventions to Mitigate Authoritarian Recursion}
\label{tab:policy-actions}
\begin{tabular}{p{0.25\textwidth} p{0.65\textwidth}}
\toprule
\textbf{Intervention} & \textbf{Description and Domain Application} \\
\midrule
\textbf{Human-in-the-loop Mandate} & Require human review before automated decisions in high-stakes domains. In education, this includes exam invalidation reviews; in warfare, it requires human authorization before lethal action. \\
\textbf{Algorithmic Explainability} & Mandate transparent audit trails for recommender systems and autonomous classifiers. Platforms must disclose signal weighting and feedback influence (e.g., YouTube’s recommendation logic). \\
\textbf{Right to Opacity} & Allow users—especially students—to opt out of invasive biometric tracking (gaze, keystroke, audio) without penalty, ensuring equitable alternatives. \\
\textbf{De-Recursive Governance} & Break feedback loops by enforcing reset intervals, diversity in training data, and limits on self-reinforcing thresholds (e.g., Harpy drone heuristics, Proctorio alert tuning). \\
\textbf{Fictional Ethics Integration} & Institutionalize the use of speculative fiction in ethics training and policy development. Narratives like \textit{Gattaca}, \textit{Men Against Fire}, and \textit{Minority Report} surface ethical blind spots not visible in technical models. \\
\bottomrule
\end{tabular}
\end{table}

\section{Conclusion}
\label{sec:conclusion}

Artificial intelligence today functions not merely as a tool of automation, but as an infrastructure of governance. Across education, warfare, and digital discourse, AI systems do not simply respond to human needs—they preempt, structure, and frequently constrain them through recursive architectures of classification, surveillance, and optimization. These infrastructures embed institutional logics into technical processes, displacing judgment and rendering control ambient yet unaccountable.

This paper has introduced the concept of \textit{authoritarian recursion} to capture how such systems consolidate normative power through feedback loops. What begins as abstraction—risk scoring, deviance flagging, content curation—quickly becomes operationalized as truth, driving future decisions with self-reinforcing certainty. The result is not a technical failure but a political one: legitimacy is ceded to opaque processes, while the possibilities for resistance or deviation shrink.

While existing ethical frameworks such as fairness, accountability, and transparency (FAccT) remain necessary, they are insufficient when reduced to checklists or retrofitted after deployment. True accountability cannot be automated; it requires the institutionalization of refusal, friction, and public oversight. Likewise, relational ethics must confront not only interpersonal dynamics but the structural asymmetries imposed by AI proxies that mediate, distort, or foreclose human interaction.

To counter these trends, this paper has proposed a range of policy interventions—outlined in Table~\ref{tab:policy-actions}—that include mandatory human-in-the-loop mechanisms, public explainability mandates, the right to opacity in educational surveillance, and the integration of speculative fiction as an ethical diagnostic. These are not technical patches but governance primitives: ways of embedding dissent, transparency, and imaginative critique into the architecture of AI itself.

Ultimately, resisting authoritarian recursion requires treating governance not as an afterthought to technical development, but as its precondition. AI governance must become a domain of democratic experimentation—a space where competing visions of justice, legitimacy, and futurity are not foreclosed by prediction, but contested in public.

\section*{AI Assistance Disclosure}

Large Language Models (LLMs), including DeepSeek-V3 and GPT-4o, were used solely for grammar correction, syntax refinement, and minor language polishing. No AI assistance was employed in generating original content, conceptual arguments, or data analysis. The author assumes full responsibility for the work's analytical integrity and originality.

\section*{Declarations}

\noindent
\textbf{Funding Declaration} \\
The author received no specific grant from any funding agency in the public, commercial, or not-for-profit sectors.

\vspace{1em}
\noindent
\textbf{Ethics Declaration} \\
This research did not involve human participants, animals, or sensitive personal data, and therefore did not require ethical approval.

\section*{Acknowledgements}

The author acknowledges Muzaffer Adak, Deniz Koca, Hatice Zor Oguz, and Ekin Ergün for valuable insights and discussions during the preparation of this manuscript.

\bibliography{references}

\begin{thebibliography}{}

\bibitem [\protect \citeauthoryear {%
Andrejevic%
}{%
Andrejevic%
}{%
{\protect \APACyear {2022}}%
}]{%
andrejevic2022automated}
\APACinsertmetastar {%
andrejevic2022automated}%
\begin{APACrefauthors}%
Andrejevic, M.%
\end{APACrefauthors}%
\unskip\
\newblock
\APACrefYear{2022}.
\newblock
\APACrefbtitle {Automated Media} {Automated media}.
\newblock
\APACaddressPublisher{}{Routledge}.
\newblock
\begin{APACrefDOI} \doi{10.4324/9780429242595} \end{APACrefDOI}
\PrintBackRefs{\CurrentBib}

\bibitem [\protect \citeauthoryear {%
Asaro%
}{%
Asaro%
}{%
{\protect \APACyear {2012}}%
}]{%
asaro2012weapons}
\APACinsertmetastar {%
asaro2012weapons}%
\begin{APACrefauthors}%
Asaro, P\BPBI M.%
\end{APACrefauthors}%
\unskip\
\newblock
\APACrefYearMonthDay{2012}{}{}.
\newblock
{\BBOQ}\APACrefatitle {On Banning Autonomous Weapon Systems: Human Rights, Automation, and the Dehumanization of Lethal Decision-Making} {On banning autonomous weapon systems: Human rights, automation, and the dehumanization of lethal decision-making}.{\BBCQ}
\newblock
\APACjournalVolNumPages{International Review of the Red Cross}{94}{886}{687--709}.
\newblock
\begin{APACrefDOI} \doi{10.1017/S1816383112000768} \end{APACrefDOI}
\PrintBackRefs{\CurrentBib}

\bibitem [\protect \citeauthoryear {%
Bahrami%
}{%
Bahrami%
}{%
{\protect \APACyear {2025}}%
}]{%
bahrami2025algemony}
\APACinsertmetastar {%
bahrami2025algemony}%
\begin{APACrefauthors}%
Bahrami, N.%
\end{APACrefauthors}%
\unskip\
\newblock
\APACrefYearMonthDay{2025}{}{}.
\newblock
{\BBOQ}\APACrefatitle {Algemony: Power Dynamics, Dominant Narratives, and Colonisation} {Algemony: Power dynamics, dominant narratives, and colonisation}.{\BBCQ}
\newblock
\APACjournalVolNumPages{AI and Ethics}{}{}{}.
\newblock
\begin{APACrefURL} \url{https://doi.org/10.1007/s43681-025-00734-4} \end{APACrefURL}
\newblock
\begin{APACrefDOI} \doi{10.1007/s43681-025-00734-4} \end{APACrefDOI}
\PrintBackRefs{\CurrentBib}

\bibitem [\protect \citeauthoryear {%
Berardi%
}{%
Berardi%
}{%
{\protect \APACyear {2015}}%
}]{%
berardi2015and}
\APACinsertmetastar {%
berardi2015and}%
\begin{APACrefauthors}%
Berardi, F.%
\end{APACrefauthors}%
\unskip\
\newblock
\APACrefYear{2015}.
\newblock
\APACrefbtitle {And: Phenomenology of the End} {And: Phenomenology of the end}.
\newblock
\APACaddressPublisher{}{MIT Press}.
\PrintBackRefs{\CurrentBib}

\bibitem [\protect \citeauthoryear {%
Black%
}{%
Black%
}{%
{\protect \APACyear {2001}}%
}]{%
black2001ibm}
\APACinsertmetastar {%
black2001ibm}%
\begin{APACrefauthors}%
Black, E.%
\end{APACrefauthors}%
\unskip\
\newblock
\APACrefYear{2001}.
\newblock
\APACrefbtitle {IBM and the Holocaust: The Strategic Alliance Between Nazi Germany and America's Most Powerful Corporation} {Ibm and the holocaust: The strategic alliance between nazi germany and america's most powerful corporation}.
\newblock
\APACaddressPublisher{}{Crown}.
\PrintBackRefs{\CurrentBib}

\bibitem [\protect \citeauthoryear {%
Bode%
\ \BBA {} Watts%
}{%
Bode%
\ \BBA {} Watts%
}{%
{\protect \APACyear {2023}}%
}]{%
bode2023loitering}
\APACinsertmetastar {%
bode2023loitering}%
\begin{APACrefauthors}%
Bode, I.%
\BCBT {}\ \BBA {} Watts, T\BPBI F\BPBI A.%
\end{APACrefauthors}%
\unskip\
\newblock
\APACrefYear{2023}.
\newblock
\APACrefbtitle {Loitering Munitions and Unpredictability: Autonomy in Weapon Systems and Challenges to Human Control} {Loitering munitions and unpredictability: Autonomy in weapon systems and challenges to human control}.
\newblock
\APACaddressPublisher{Odense / London}{Center for War Studies}.
\newblock
\begin{APACrefURL} \url{https://pure.royalholloway.ac.uk/files/55721928/Loitering-Munitions-Unpredictability-WEB.pdf} \end{APACrefURL}
\newblock
\APACrefnote{Open‑access report, accessed July 2025}
\PrintBackRefs{\CurrentBib}

\bibitem [\protect \citeauthoryear {%
Cave%
\ \BBA {} Dihal%
}{%
Cave%
\ \BBA {} Dihal%
}{%
{\protect \APACyear {2020}}%
}]{%
cave2020whiteness}
\APACinsertmetastar {%
cave2020whiteness}%
\begin{APACrefauthors}%
Cave, S.%
\BCBT {}\ \BBA {} Dihal, K.%
\end{APACrefauthors}%
\unskip\
\newblock
\APACrefYearMonthDay{2020}{}{}.
\newblock
{\BBOQ}\APACrefatitle {The Whiteness of AI} {The whiteness of ai}.{\BBCQ}
\newblock
\APACjournalVolNumPages{Philosophy \& Technology}{33}{4}{685--703}.
\newblock
\begin{APACrefDOI} \doi{10.1007/s13347-020-00403-5} \end{APACrefDOI}
\PrintBackRefs{\CurrentBib}

\bibitem [\protect \citeauthoryear {%
Cave%
, Dihal%
\BCBL {}\ \BBA {} Dillon%
}{%
Cave%
\ \protect \BOthers {.}}{%
{\protect \APACyear {2019}}%
}]{%
cave2019ai}
\APACinsertmetastar {%
cave2019ai}%
\begin{APACrefauthors}%
Cave, S.%
, Dihal, K.%
\BCBL {}\ \BBA {} Dillon, S.%
\end{APACrefauthors}%
\unskip\
\newblock
\APACrefYearMonthDay{2019}{}{}.
\newblock
{\BBOQ}\APACrefatitle {AI Narratives: A History of Imaginative Thinking about Intelligent Machines} {Ai narratives: A history of imaginative thinking about intelligent machines}.{\BBCQ}
\newblock
\APACjournalVolNumPages{Science Fiction Studies}{46}{3}{473--495}.
\newblock
\begin{APACrefDOI} \doi{10.5621/sciefictstud.46.3.0473} \end{APACrefDOI}
\PrintBackRefs{\CurrentBib}

\bibitem [\protect \citeauthoryear {%
Coeckelbergh%
}{%
Coeckelbergh%
}{%
{\protect \APACyear {2020}}%
}]{%
coeckelbergh2020aiethics}
\APACinsertmetastar {%
coeckelbergh2020aiethics}%
\begin{APACrefauthors}%
Coeckelbergh, M.%
\end{APACrefauthors}%
\unskip\
\newblock
\APACrefYear{2020}.
\newblock
\APACrefbtitle {AI Ethics} {Ai ethics}.
\newblock
\APACaddressPublisher{}{MIT Press}.
\newblock
\begin{APACrefDOI} \doi{10.7551/mitpress/12294.001.0001} \end{APACrefDOI}
\PrintBackRefs{\CurrentBib}

\bibitem [\protect \citeauthoryear {%
Cottom%
}{%
Cottom%
}{%
{\protect \APACyear {2020}}%
}]{%
Cottom2020}
\APACinsertmetastar {%
Cottom2020}%
\begin{APACrefauthors}%
Cottom, T\BPBI M.%
\end{APACrefauthors}%
\unskip\
\newblock
\APACrefYearMonthDay{2020}{}{}.
\newblock
{\BBOQ}\APACrefatitle {Where Platform Capitalism and Racial Capitalism Meet: The Sociology of Race and Racism in the Digital Society} {Where platform capitalism and racial capitalism meet: The sociology of race and racism in the digital society}.{\BBCQ}
\newblock
\APACjournalVolNumPages{Sociology of Race and Ethnicity}{6}{4}{441-449}.
\newblock
\begin{APACrefURL} \url{https://doi.org/10.1177/2332649220949473} \end{APACrefURL}
\newblock
\begin{APACrefDOI} \doi{10.1177/2332649220949473} \end{APACrefDOI}
\PrintBackRefs{\CurrentBib}

\bibitem [\protect \citeauthoryear {%
Couldry%
\ \BBA {} Mejias%
}{%
Couldry%
\ \BBA {} Mejias%
}{%
{\protect \APACyear {2019}}%
}]{%
couldry2019data}
\APACinsertmetastar {%
couldry2019data}%
\begin{APACrefauthors}%
Couldry, N.%
\BCBT {}\ \BBA {} Mejias, U\BPBI A.%
\end{APACrefauthors}%
\unskip\
\newblock
\APACrefYear{2019}.
\newblock
\APACrefbtitle {The Costs of Connection: How Data Is Colonizing Human Life and Appropriating It for Capitalism} {The costs of connection: How data is colonizing human life and appropriating it for capitalism}.
\newblock
\APACaddressPublisher{}{Stanford University Press}.
\PrintBackRefs{\CurrentBib}

\bibitem [\protect \citeauthoryear {%
Deleuze%
}{%
Deleuze%
}{%
{\protect \APACyear {1992}}%
}]{%
deleuze1992postscript}
\APACinsertmetastar {%
deleuze1992postscript}%
\begin{APACrefauthors}%
Deleuze, G.%
\end{APACrefauthors}%
\unskip\
\newblock
\APACrefYearMonthDay{1992}{}{}.
\newblock
{\BBOQ}\APACrefatitle {Postscript on the Societies of Control} {Postscript on the societies of control}.{\BBCQ}
\newblock
\APACjournalVolNumPages{October}{59}{}{3--7}.
\newblock
\begin{APACrefURL} \url{https://www.jstor.org/stable/778828} \end{APACrefURL}
\PrintBackRefs{\CurrentBib}

\bibitem [\protect \citeauthoryear {%
Dencik%
, Hintz%
, Redden%
\BCBL {}\ \BBA {} Treré%
}{%
Dencik%
\ \protect \BOthers {.}}{%
{\protect \APACyear {2019}}%
}]{%
dencik2019exploring}
\APACinsertmetastar {%
dencik2019exploring}%
\begin{APACrefauthors}%
Dencik, L.%
, Hintz, A.%
, Redden, J.%
\BCBL {}\ \BBA {} Treré, E.%
\end{APACrefauthors}%
\unskip\
\newblock
\APACrefYearMonthDay{2019}{}{}.
\newblock
{\BBOQ}\APACrefatitle {Exploring Data Justice: Conceptions, Applications and Directions} {Exploring data justice: Conceptions, applications and directions}.{\BBCQ}
\newblock
\APACjournalVolNumPages{Information, Communication \& Society}{22}{7}{873--881}.
\newblock
\begin{APACrefDOI} \doi{10.1080/1369118X.2019.1606268} \end{APACrefDOI}
\PrintBackRefs{\CurrentBib}

\bibitem [\protect \citeauthoryear {%
{European Commission}%
}{%
{European Commission}%
}{%
{\protect \APACyear {2021}}%
}]{%
eu2021aiact}
\APACinsertmetastar {%
eu2021aiact}%
\begin{APACrefauthors}%
{European Commission}.%
\end{APACrefauthors}%
\unskip\
\newblock
\APACrefYearMonthDay{2021}{April}{}.
\newblock
\APACrefbtitle {Proposal for a Regulation Laying Down Harmonised Rules on Artificial Intelligence (Artificial Intelligence Act).} {Proposal for a regulation laying down harmonised rules on artificial intelligence (artificial intelligence act).}
\newblock
\begin{APACrefURL} \url{https://eur-lex.europa.eu/legal-content/EN/TXT/?uri=CELEX%3A52021PC0206} \end{APACrefURL}
\newblock
\APACrefnote{COM(2021) 206 final}
\PrintBackRefs{\CurrentBib}

\bibitem [\protect \citeauthoryear {%
Fairclough%
}{%
Fairclough%
}{%
{\protect \APACyear {1992}}%
}]{%
fairclough1992discourse}
\APACinsertmetastar {%
fairclough1992discourse}%
\begin{APACrefauthors}%
Fairclough, N.%
\end{APACrefauthors}%
\unskip\
\newblock
\APACrefYear{1992}.
\newblock
\APACrefbtitle {Discourse and Social Change} {Discourse and social change}.
\newblock
\APACaddressPublisher{}{Polity Press}.
\PrintBackRefs{\CurrentBib}

\bibitem [\protect \citeauthoryear {%
Fisher%
}{%
Fisher%
}{%
{\protect \APACyear {2018}}%
}]{%
fisher2018youtube}
\APACinsertmetastar {%
fisher2018youtube}%
\begin{APACrefauthors}%
Fisher, M.%
\end{APACrefauthors}%
\unskip\
\newblock
\APACrefYearMonthDay{2018}{}{}.
\newblock
\APACrefbtitle {YouTube, the Great Radicalizer.} {Youtube, the great radicalizer.}
\newblock
\APAChowpublished {\textit{The New York Times}, March 10}.
\newblock
\begin{APACrefURL} \url{https://www.nytimes.com/2018/03/10/opinion/sunday/youtube-politics-radical.html} \end{APACrefURL}
\PrintBackRefs{\CurrentBib}

\bibitem [\protect \citeauthoryear {%
Foucault%
}{%
Foucault%
}{%
{\protect \APACyear {1995}}%
}]{%
foucault1995discipline}
\APACinsertmetastar {%
foucault1995discipline}%
\begin{APACrefauthors}%
Foucault, M.%
\end{APACrefauthors}%
\unskip\
\newblock
\APACrefYear{1995}.
\newblock
\APACrefbtitle {Discipline and Punish: The Birth of the Prison} {Discipline and punish: The birth of the prison}.
\newblock
\APACaddressPublisher{}{Vintage Books}.
\PrintBackRefs{\CurrentBib}

\bibitem [\protect \citeauthoryear {%
Gillespie%
}{%
Gillespie%
}{%
{\protect \APACyear {2018}}%
}]{%
gillespie2018custodians}
\APACinsertmetastar {%
gillespie2018custodians}%
\begin{APACrefauthors}%
Gillespie, T.%
\end{APACrefauthors}%
\unskip\
\newblock
\APACrefYear{2018}.
\newblock
\APACrefbtitle {Custodians of the Internet: Platforms, Content Moderation, and the Hidden Decisions That Shape Social Media} {Custodians of the internet: Platforms, content moderation, and the hidden decisions that shape social media}.
\newblock
\APACaddressPublisher{}{Yale University Press}.
\PrintBackRefs{\CurrentBib}

\bibitem [\protect \citeauthoryear {%
Gilliard%
\ \BBA {} Selwyn%
}{%
Gilliard%
\ \BBA {} Selwyn%
}{%
{\protect \APACyear {2023}}%
}]{%
gilliard2023automated}
\APACinsertmetastar {%
gilliard2023automated}%
\begin{APACrefauthors}%
Gilliard, C.%
\BCBT {}\ \BBA {} Selwyn, N.%
\end{APACrefauthors}%
\unskip\
\newblock
\APACrefYearMonthDay{2023}{}{}.
\newblock
{\BBOQ}\APACrefatitle {Automated Surveillance in Education} {Automated surveillance in education}.{\BBCQ}
\newblock
\APACjournalVolNumPages{Postdigital Science and Education}{5}{1}{195--205}.
\newblock
\begin{APACrefURL} \url{https://doi.org/10.1007/s42438-022-00295-3} \end{APACrefURL}
\newblock
\begin{APACrefDOI} \doi{10.1007/s42438-022-00295-3} \end{APACrefDOI}
\PrintBackRefs{\CurrentBib}

\bibitem [\protect \citeauthoryear {%
{Google Research}%
}{%
{Google Research}%
}{%
{\protect \APACyear {2022}}%
}]{%
google2022recsys}
\APACinsertmetastar {%
google2022recsys}%
\begin{APACrefauthors}%
{Google Research}.%
\end{APACrefauthors}%
\unskip\
\newblock
\APACrefYearMonthDay{2022}{}{}.
\newblock
\APACrefbtitle {TensorFlow Recommenders: Building Scalable Recommender Systems.} {Tensorflow recommenders: Building scalable recommender systems.}
\newblock
\begin{APACrefURL} \url{https://www.tensorflow.org/recommenders} \end{APACrefURL}
\newblock
\APACrefnote{Technical framework for recommendation modeling}
\PrintBackRefs{\CurrentBib}

\bibitem [\protect \citeauthoryear {%
Hanna%
\ \BBA {} Kazim%
}{%
Hanna%
\ \BBA {} Kazim%
}{%
{\protect \APACyear {2021}}%
}]{%
Hanna2021}
\APACinsertmetastar {%
Hanna2021}%
\begin{APACrefauthors}%
Hanna, R.%
\BCBT {}\ \BBA {} Kazim, E.%
\end{APACrefauthors}%
\unskip\
\newblock
\APACrefYearMonthDay{2021}{}{}.
\newblock
{\BBOQ}\APACrefatitle {Philosophical Foundations for Digital Ethics and {AI} Ethics: A Dignitarian Approach} {Philosophical foundations for digital ethics and {AI} ethics: A dignitarian approach}.{\BBCQ}
\newblock
\APACjournalVolNumPages{AI and Ethics}{1}{4}{405--423}.
\newblock
\begin{APACrefDOI} \doi{10.1007/s43681-021-00040-9} \end{APACrefDOI}
\PrintBackRefs{\CurrentBib}

\bibitem [\protect \citeauthoryear {%
{IEEE Global Initiative on Ethics of Autonomous and Intelligent Systems}%
}{%
{IEEE Global Initiative on Ethics of Autonomous and Intelligent Systems}%
}{%
{\protect \APACyear {2019}}%
}]{%
ieee2019ethicallyaligned}
\APACinsertmetastar {%
ieee2019ethicallyaligned}%
\begin{APACrefauthors}%
{IEEE Global Initiative on Ethics of Autonomous and Intelligent Systems}.%
\end{APACrefauthors}%
\unskip\
\newblock
\APACrefYearMonthDay{2019}{}{}.
\newblock
\APACrefbtitle {Ethically Aligned Design: A Vision for Prioritizing Human Well-being with Autonomous and Intelligent Systems} {Ethically aligned design: A vision for prioritizing human well-being with autonomous and intelligent systems}\ (\PrintOrdinal{1st}\ \BEd).
\newblock
\begin{APACrefURL} \url{https://standards.ieee.org/industry-connections/ec/autonomous-systems.html} \end{APACrefURL}
\PrintBackRefs{\CurrentBib}

\bibitem [\protect \citeauthoryear {%
Lupton%
}{%
Lupton%
}{%
{\protect \APACyear {2017}}%
}]{%
lupton2017feeling}
\APACinsertmetastar {%
lupton2017feeling}%
\begin{APACrefauthors}%
Lupton, D.%
\end{APACrefauthors}%
\unskip\
\newblock
\APACrefYearMonthDay{2017}{}{}.
\newblock
{\BBOQ}\APACrefatitle {Feeling your data: Touch and making sense of personal digital data} {Feeling your data: Touch and making sense of personal digital data}.{\BBCQ}
\newblock
\APACjournalVolNumPages{New Media \& Society}{19}{10}{1599-1614}.
\newblock
\begin{APACrefURL} \url{https://doi.org/10.1177/1461444817717515} \end{APACrefURL}
\newblock
\begin{APACrefDOI} \doi{10.1177/1461444817717515} \end{APACrefDOI}
\PrintBackRefs{\CurrentBib}

\bibitem [\protect \citeauthoryear {%
Marsili%
}{%
Marsili%
}{%
{\protect \APACyear {2024}}%
}]{%
marsili2024}
\APACinsertmetastar {%
marsili2024}%
\begin{APACrefauthors}%
Marsili, M.%
\end{APACrefauthors}%
\unskip\
\newblock
\APACrefYearMonthDay{2024}{Oct.}{}.
\newblock
{\BBOQ}\APACrefatitle {Lethal Autonomous Weapon Systems: Ethical Dilemmas and Legal Compliance in the Era of Military Disruptive Technologies} {Lethal autonomous weapon systems: Ethical dilemmas and legal compliance in the era of military disruptive technologies}.{\BBCQ}
\newblock
\APACjournalVolNumPages{International Journal of Robotics and Automation Technology}{11}{}{63–68}.
\newblock
\begin{APACrefURL} \url{https://zealpress.com/jms/index.php/ijrat/article/view/599} \end{APACrefURL}
\newblock
\begin{APACrefDOI} \doi{10.31875/2409-9694.2024.11.05} \end{APACrefDOI}
\PrintBackRefs{\CurrentBib}

\bibitem [\protect \citeauthoryear {%
Mittelstadt%
, Allo%
, Taddeo%
, Wachter%
\BCBL {}\ \BBA {} Floridi%
}{%
Mittelstadt%
\ \protect \BOthers {.}}{%
{\protect \APACyear {2016}}%
}]{%
mittelstadt2016ethics}
\APACinsertmetastar {%
mittelstadt2016ethics}%
\begin{APACrefauthors}%
Mittelstadt, B.%
, Allo, P.%
, Taddeo, M.%
, Wachter, S.%
\BCBL {}\ \BBA {} Floridi, L.%
\end{APACrefauthors}%
\unskip\
\newblock
\APACrefYearMonthDay{2016}{}{}.
\newblock
{\BBOQ}\APACrefatitle {The ethics of algorithms: Mapping the debate} {The ethics of algorithms: Mapping the debate}.{\BBCQ}
\newblock
\APACjournalVolNumPages{Big Data \& Society}{3}{2}{1--21}.
\newblock
\begin{APACrefURL} \url{https://doi.org/10.1177/2053951716679679} \end{APACrefURL}
\newblock
\begin{APACrefDOI} \doi{10.1177/2053951716679679} \end{APACrefDOI}
\PrintBackRefs{\CurrentBib}

\bibitem [\protect \citeauthoryear {%
{Mozilla Foundation}%
}{%
{Mozilla Foundation}%
}{%
{\protect \APACyear {2021}}%
}]{%
mozilla2021regrets}
\APACinsertmetastar {%
mozilla2021regrets}%
\begin{APACrefauthors}%
{Mozilla Foundation}.%
\end{APACrefauthors}%
\unskip\
\newblock
\APACrefYearMonthDay{2021}{}{}.
\newblock
\APACrefbtitle {YouTube Regrets: A Crowdsourced Investigation into Algorithmic Harm.} {Youtube regrets: A crowdsourced investigation into algorithmic harm.}
\newblock
\begin{APACrefURL} \url{https://foundation.mozilla.org/en/youtube/regrets/} \end{APACrefURL}
\newblock
\APACrefnote{Mozilla RegretsReporter Project, retrieved July 2025}
\PrintBackRefs{\CurrentBib}

\bibitem [\protect \citeauthoryear {%
Noble%
}{%
Noble%
}{%
{\protect \APACyear {2018}}%
}]{%
noble2018algorithms}
\APACinsertmetastar {%
noble2018algorithms}%
\begin{APACrefauthors}%
Noble, S\BPBI U.%
\end{APACrefauthors}%
\unskip\
\newblock
\APACrefYear{2018}.
\newblock
\APACrefbtitle {Algorithms of Oppression: How Search Engines Reinforce Racism} {Algorithms of oppression: How search engines reinforce racism}.
\newblock
\APACaddressPublisher{}{NYU Press}.
\PrintBackRefs{\CurrentBib}

\bibitem [\protect \citeauthoryear {%
{Organisation for Economic Co-operation and Development}%
}{%
{Organisation for Economic Co-operation and Development}%
}{%
{\protect \APACyear {2019}}%
}]{%
oecd2019principles}
\APACinsertmetastar {%
oecd2019principles}%
\begin{APACrefauthors}%
{Organisation for Economic Co-operation and Development}.%
\end{APACrefauthors}%
\unskip\
\newblock
\APACrefYearMonthDay{2019}{}{}.
\newblock
\APACrefbtitle {OECD Principles on Artificial Intelligence.} {Oecd principles on artificial intelligence.}
\newblock
\begin{APACrefURL} \url{https://legalinstruments.oecd.org/en/instruments/OECD-LEGAL-0449} \end{APACrefURL}
\newblock
\APACrefnote{Adopted by OECD Council, May 2019}
\PrintBackRefs{\CurrentBib}

\bibitem [\protect \citeauthoryear {%
Paul%
}{%
Paul%
}{%
{\protect \APACyear {2021}}%
}]{%
paul2021proctorio}
\APACinsertmetastar {%
paul2021proctorio}%
\begin{APACrefauthors}%
Paul, K.%
\end{APACrefauthors}%
\unskip\
\newblock
\APACrefYearMonthDay{2021}{March}{}.
\newblock
\APACrefbtitle {Proctorio Is Using Racist Algorithms to Detect Faces.} {Proctorio is using racist algorithms to detect faces.}
\newblock
\begin{APACrefURL} \url{https://www.vice.com/en/article/proctorio-is-using-racist-algorithms-to-detect-faces/} \end{APACrefURL}
\newblock
\APACrefnote{Vice News, retrieved July 2025}
\PrintBackRefs{\CurrentBib}

\bibitem [\protect \citeauthoryear {%
{Proctorio}%
}{%
{Proctorio}%
}{%
{\protect \APACyear {2023}}%
}]{%
proctorio2023docs}
\APACinsertmetastar {%
proctorio2023docs}%
\begin{APACrefauthors}%
{Proctorio}.%
\end{APACrefauthors}%
\unskip\
\newblock
\APACrefYearMonthDay{2023}{}{}.
\newblock
\APACrefbtitle {Privacy and Security Practices: Transparency and Technical Safeguards.} {Privacy and security practices: Transparency and technical safeguards.}
\newblock
\begin{APACrefURL} \url{https://proctorio.com/privacy} \end{APACrefURL}
\newblock
\APACrefnote{Accessed July 2025}
\PrintBackRefs{\CurrentBib}

\bibitem [\protect \citeauthoryear {%
Raji%
\ \protect \BOthers {.}}{%
Raji%
\ \protect \BOthers {.}}{%
{\protect \APACyear {2020}}%
}]{%
raji2020closing}
\APACinsertmetastar {%
raji2020closing}%
\begin{APACrefauthors}%
Raji, I\BPBI D.%
, Dobbe, R.%
, Hutchinson, B.%
, Mitchell, M.%
, Gebru, T.%
, Barnes, J.%
\BCBL {}\ \BBA {} Raji, I.%
\end{APACrefauthors}%
\unskip\
\newblock
\APACrefYearMonthDay{2020}{}{}.
\newblock
{\BBOQ}\APACrefatitle {Closing the AI accountability gap: Defining an end-to-end framework for internal algorithmic auditing} {Closing the ai accountability gap: Defining an end-to-end framework for internal algorithmic auditing}.{\BBCQ}
\newblock
\APACjournalVolNumPages{Proceedings of the 2020 Conference on Fairness, Accountability, and Transparency}{}{}{33--44}.
\newblock
\begin{APACrefDOI} \doi{10.1145/3351095.3372873} \end{APACrefDOI}
\PrintBackRefs{\CurrentBib}

\bibitem [\protect \citeauthoryear {%
Roy-Stang%
\ \BBA {} Davies%
}{%
Roy-Stang%
\ \BBA {} Davies%
}{%
{\protect \APACyear {2025}}%
}]{%
Roy-Stang2025}
\APACinsertmetastar {%
Roy-Stang2025}%
\begin{APACrefauthors}%
Roy-Stang, Z.%
\BCBT {}\ \BBA {} Davies, J.%
\end{APACrefauthors}%
\unskip\
\newblock
\APACrefYearMonthDay{2025}{}{}.
\newblock
{\BBOQ}\APACrefatitle {Human Biases and Remedies in {AI} Safety and Alignment Contexts} {Human biases and remedies in {AI} safety and alignment contexts}.{\BBCQ}
\newblock
\APACjournalVolNumPages{AI and Ethics}{5}{2}{123--145}.
\newblock
\begin{APACrefURL} \url{https://doi.org/10.1007/s43681-025-00698-5} \end{APACrefURL}
\newblock
\begin{APACrefDOI} \doi{10.1007/s43681-025-00698-5} \end{APACrefDOI}
\PrintBackRefs{\CurrentBib}

\bibitem [\protect \citeauthoryear {%
Selwyn%
, Hillman%
, Bergviken~Rensfeldt%
\BCBL {}\ \BBA {} Perrotta%
}{%
Selwyn%
\ \protect \BOthers {.}}{%
{\protect \APACyear {2023}}%
}]{%
selwyn2023digital}
\APACinsertmetastar {%
selwyn2023digital}%
\begin{APACrefauthors}%
Selwyn, N.%
, Hillman, T.%
, Bergviken~Rensfeldt, A.%
\BCBL {}\ \BBA {} Perrotta, C.%
\end{APACrefauthors}%
\unskip\
\newblock
\APACrefYearMonthDay{2023}{}{}.
\newblock
{\BBOQ}\APACrefatitle {Digital Technologies and the Automation of Education—Key Questions and Concerns} {Digital technologies and the automation of education—key questions and concerns}.{\BBCQ}
\newblock
\APACjournalVolNumPages{Postdigital Science and Education}{5}{1}{15--24}.
\newblock
\begin{APACrefURL} \url{https://doi.org/10.1007/s42438-021-00263-3} \end{APACrefURL}
\newblock
\begin{APACrefDOI} \doi{10.1007/s42438-021-00263-3} \end{APACrefDOI}
\PrintBackRefs{\CurrentBib}

\bibitem [\protect \citeauthoryear {%
Tufekci%
}{%
Tufekci%
}{%
{\protect \APACyear {2015}}%
}]{%
tufekci2015algorithmic}
\APACinsertmetastar {%
tufekci2015algorithmic}%
\begin{APACrefauthors}%
Tufekci, Z.%
\end{APACrefauthors}%
\unskip\
\newblock
\APACrefYearMonthDay{2015}{}{}.
\newblock
{\BBOQ}\APACrefatitle {Algorithmic harms beyond Facebook and Google: Emergent challenges of computational agency} {Algorithmic harms beyond facebook and google: Emergent challenges of computational agency}.{\BBCQ}
\newblock
\APACjournalVolNumPages{Colorado Technology Law Journal}{13}{203}{203--218}.
\PrintBackRefs{\CurrentBib}

\bibitem [\protect \citeauthoryear {%
{UNESCO}%
}{%
{UNESCO}%
}{%
{\protect \APACyear {2021}}%
}]{%
unesco2021ethics}
\APACinsertmetastar {%
unesco2021ethics}%
\begin{APACrefauthors}%
{UNESCO}.%
\end{APACrefauthors}%
\unskip\
\newblock
\APACrefYearMonthDay{2021}{}{}.
\newblock
\APACrefbtitle {Recommendation on the Ethics of Artificial Intelligence.} {Recommendation on the ethics of artificial intelligence.}
\newblock
\begin{APACrefURL} \url{https://unesdoc.unesco.org/ark:/48223/pf0000381137} \end{APACrefURL}
\newblock
\APACrefnote{Adopted at the 41st session of the General Conference}
\PrintBackRefs{\CurrentBib}

\bibitem [\protect \citeauthoryear {%
{U.S. Department of Defense}%
}{%
{U.S. Department of Defense}%
}{%
{\protect \APACyear {2020}}%
}]{%
dod2020principles}
\APACinsertmetastar {%
dod2020principles}%
\begin{APACrefauthors}%
{U.S. Department of Defense}.%
\end{APACrefauthors}%
\unskip\
\newblock
\APACrefYearMonthDay{2020}{February}{}.
\newblock
\APACrefbtitle {Ethical Principles for Artificial Intelligence.} {Ethical principles for artificial intelligence.}
\newblock
\begin{APACrefURL} \url{https://www.defense.gov/News/Releases/Release/Article/2091996/dod-adopts-ethical-principles-for-artificial-intelligence/} \end{APACrefURL}
\newblock
\APACrefnote{Published by the Defense Innovation Board}
\PrintBackRefs{\CurrentBib}

\bibitem [\protect \citeauthoryear {%
van Dijck%
, Poell%
\BCBL {}\ \BBA {} de Waal%
}{%
van Dijck%
\ \protect \BOthers {.}}{%
{\protect \APACyear {2018}}%
}]{%
vanDijck2018platform}
\APACinsertmetastar {%
vanDijck2018platform}%
\begin{APACrefauthors}%
van Dijck, J.%
, Poell, T.%
\BCBL {}\ \BBA {} de Waal, M.%
\end{APACrefauthors}%
\unskip\
\newblock
\APACrefYear{2018}.
\newblock
\APACrefbtitle {The Platform Society: Public Values in a Connective World} {The platform society: Public values in a connective world}.
\newblock
\APACaddressPublisher{}{Oxford University Press}.
\PrintBackRefs{\CurrentBib}

\bibitem [\protect \citeauthoryear {%
Vosoughi%
, Roy%
\BCBL {}\ \BBA {} Aral%
}{%
Vosoughi%
\ \protect \BOthers {.}}{%
{\protect \APACyear {2018}}%
}]{%
vosoughi2018spread}
\APACinsertmetastar {%
vosoughi2018spread}%
\begin{APACrefauthors}%
Vosoughi, S.%
, Roy, D.%
\BCBL {}\ \BBA {} Aral, S.%
\end{APACrefauthors}%
\unskip\
\newblock
\APACrefYearMonthDay{2018}{}{}.
\newblock
{\BBOQ}\APACrefatitle {The spread of true and false news online} {The spread of true and false news online}.{\BBCQ}
\newblock
\APACjournalVolNumPages{Science}{359}{6380}{1146--1151}.
\newblock
\begin{APACrefDOI} \doi{10.1126/science.aap9559} \end{APACrefDOI}
\PrintBackRefs{\CurrentBib}

\bibitem [\protect \citeauthoryear {%
Williams%
, Brooks%
\BCBL {}\ \BBA {} Shapiro%
}{%
Williams%
\ \protect \BOthers {.}}{%
{\protect \APACyear {2022}}%
}]{%
williams2022racial}
\APACinsertmetastar {%
williams2022racial}%
\begin{APACrefauthors}%
Williams, C\BPBI M.%
, Brooks, J\BPBI L.%
\BCBL {}\ \BBA {} Shapiro, A\BPBI R.%
\end{APACrefauthors}%
\unskip\
\newblock
\APACrefYearMonthDay{2022}{}{}.
\newblock
{\BBOQ}\APACrefatitle {Racial, Skin Tone, and Sex Disparities in Automated Proctoring Software} {Racial, skin tone, and sex disparities in automated proctoring software}.{\BBCQ}
\newblock
\APACjournalVolNumPages{Frontiers in Education}{7}{}{881449}.
\newblock
\begin{APACrefURL} \url{https://www.frontiersin.org/articles/10.3389/feduc.2022.881449/full} \end{APACrefURL}
\newblock
\begin{APACrefDOI} \doi{10.3389/feduc.2022.881449} \end{APACrefDOI}
\PrintBackRefs{\CurrentBib}

\bibitem [\protect \citeauthoryear {%
Wylie%
}{%
Wylie%
}{%
{\protect \APACyear {2019}}%
}]{%
wylie2019mindfck}
\APACinsertmetastar {%
wylie2019mindfck}%
\begin{APACrefauthors}%
Wylie, C.%
\end{APACrefauthors}%
\unskip\
\newblock
\APACrefYear{2019}.
\newblock
\APACrefbtitle {Mindf*ck: Cambridge Analytica and the Plot to Break America} {Mindf*ck: Cambridge analytica and the plot to break america}.
\newblock
\APACaddressPublisher{}{Random House}.
\PrintBackRefs{\CurrentBib}

\bibitem [\protect \citeauthoryear {%
Zuboff%
}{%
Zuboff%
}{%
{\protect \APACyear {2019}}%
}]{%
zuboff2019age}
\APACinsertmetastar {%
zuboff2019age}%
\begin{APACrefauthors}%
Zuboff, S.%
\end{APACrefauthors}%
\unskip\
\newblock
\APACrefYear{2019}.
\newblock
\APACrefbtitle {The Age of Surveillance Capitalism: The Fight for a Human Future at the New Frontier of Power} {The age of surveillance capitalism: The fight for a human future at the new frontier of power}.
\newblock
\APACaddressPublisher{}{PublicAffairs}.
\PrintBackRefs{\CurrentBib}

\end{thebibliography}
\end{document}